\documentclass[aps,showpacs,reprint]{revtex4-1}  
\usepackage{epsf}
\usepackage{bm}
\usepackage{epsfig}
\usepackage{color}
\usepackage{graphicx}
\begin{document}

\preprint{APS/123-QED}  

\title{Effect of C$_{60}$ giant resonance on the photoabsorption of   
encaged atoms} 
\author{Zhifan Chen}  
\author{Alfred Z Msezane}
\affiliation{
Department of Physics and CTSPS, Clark Atlanta University, Atlanta, GA 30314 USA}
\date{\today}
 
\begin{abstract} 
    The absolute differential oscillator strengths (DOS's) for the photoabsorption of the  
Ne, Ar, and Xe atoms encapsulated in the C$_{60}$ have been evaluated using the 
time-dependent-density-functional-theory, which solves the quantum Liouvillian equation
with the Lanczos chain method. The calculations are performed in the energy regions 
both inside and outside
the C$_{60}$ giant resonance. The photoabsorption spectra of the atoms encaged in the C$_{60}$ 
demonstrate strong oscillations inside the energy range of the C$_{60}$ giant resonance.
This type of oscillation cannot be explained by the confinement resonance, but is due  
to the energy transfer from the C$_{60}$ valence electrons to the photoelectron 
through the intershell coupling. 
\end{abstract}
\maketitle

\section{Introduction}
Recently a method to evaluate the absolute differential oscillator strengths (DOS's)
for the photoabsorption of atoms encapsulated inside a  
fullerene has been developed \cite{r1,r2}. This method takes three steps to 
calculate the photoabsorption spectra of the fullerene and the 
endohedral fullerene separately. Firstly, the structures of 
the fullerene and endohedral fullerene are optimized \cite{c9}. Secondly, the ground state eigenvalues and
eigenvectors are created by solving the Kohn-Sham equation self-consistently using 
supercell \cite{c10} and a plane wave approach \cite{c11}. Thirdly, the linear 
response of the system to the perturbation 
by an external electric field is described by the quantum Liouvillian equation \cite{c13}. 
The photoabsorption spectra are evaluated using  
the time-dependent-density-functional-theory (TDDFT) 
with the Liouville-Lanczos approach \cite{c14,c15}.  
Finally, the absolute DOS's for the photoabsorption of an atom 
encapsulated inside a fullerene can be calculated by subtracting the DOS's of the fullerene 
from the correspondent DOS's of the endohedral fullerene at the same photon energy.
 
This method has been successfully used to study the photoabsorption spectra of 
the Xe@C$_{60}$ [1] and Sc$_3$N@C$_{80}$ [2] molecules. These examples 
demonstrate the great advantage of the method. The method can be used to study the atoms located at the 
center of the fullerene [1] and the off-center positions as well [2].  
It also allows us to evaluate the spectrum in a broad energy region.

In the photoionization studies of endohedral fullerenes the most interesting  
problem is probably
the Xe 4$d$ giant resonance of the Xe@C$_{60}$ molecule \cite{r3,r4,amu,r5,r6,d16}. 
The theoretical calculations that used model potentials \cite{r4,r6} and the TDDFT 
with jellium approximation \cite{r5} ignored    
the effect of the C$_{60}$ giant resonance because, fortunately
the Xe 4$d$ giant resonance is found at the energy region far from the C$_{60}$ 
resonance. 
Amusia and Balenkov \cite{r7} and Madjet {\it et al} \cite{r8} have studied
the C$_{60}$ resonance effects on the photoionization cross sections of the 
Xe 5$s$ and Ar 3$p$ electrons encaged in the C$_{60}$.
However, the photoabsorption spectra
of the Ar and Xe atoms encapsulated in the C$_{60}$ is still unknown.
In this paper we have evaluated the absolute DOS's for the photoabsorption of 
the Ne, Ar, and Xe atoms encapsulated in the C$_{60}$ in 
the regions both inside and outside the C$_{60}$ giant resonance. The 
results demonstrate 
a strong oscillation and a significant increase of the DOS's 
in the C$_{60}$ resonance region for all these atoms.

\section{Method}
     As stated in the Introduction the method of calculation used here takes three 
steps to evaluate the photoabsorption spectra of the fullerene 
and endohedral fullerene.  
In the first step we utilize the DMol$_3$ 
software package \cite{c9} to determine the optimized structure of the fullerene. 
Geometry optimization of the C$_{60}$ was performed using the generalized gradient
approximation (GGA) to the density-functional-theory (DFT) \cite{c10}, with 
Perdew-Burke-Ernzerhof (PBE)
exchange-correlation functional \cite{r10} along with all electrons double numerical plus 
polarization (DNP) basis sets and dispersion correction as implemented in the DMol$_3$ 
package \cite{c9}. The optimization of atomic positions proceeded until the change in 
energy was less than 5$\times$10$^{-4}$ eV and the forces were less than 0.01 eV/{\AA}.
The optimized structure was then introduced into a supercell of 18\AA. 
The Kohn-Sham equation
was solved self-consistently to create the ground state eigenstates and eigenvalues 
for the total of 240 electrons and 120 states using the plane wave
approach \cite{c11}. An ultrasoft pseudopotential using Rappe-Rabe-Kaxiras-Joannopoulos (RRKJ) 
pseudization algorithms \cite{r12}, which replace atomic orbitals in the 
core region with smooth nodeless pseudo-orbitals, has been employed in the calculation.
The kinetic energy cutoff 
of 408 eV for the wave function and 2448 eV 
for the densities and potentials were employed in a standard ground state DFT 
calculation \cite{c11}. 
The linear response of the ground state to an external perturbation by an electric 
field was described by the quantum Liouvillian equation  
\cite{c13, c14}(atomic units are used throughout, unless stated otherwise): 

\begin{equation}
i{d \rho \over dt}=[{H_{KS}}(t), \rho (t)],  
\end{equation}
where $H_{KS}(t)$ is time dependent Kohn-Sham (KS) Hamiltonian and $\rho (t)$ is 
one-electron KS density matrix.
The KS Hamiltonian can be written as
\begin{equation}
{H_{KS}}(t) =-{1\over 2} \nabla^2 + v_{ext}({\bf r},t) + v_{Hxc} ({\bf r}, t)  
\end{equation}
where $v_{ext}({\bf r},t)$ is the external potential
and $v_{Hxc}({\bf r},t)$ is the time dependent Hartree potential plus exchange-correlation potential.

Linearization of Eq. (1) with respect to the external perturbation leads to:
\begin{equation} 
i{d {\rho'} \over dt}= [{H^{GS}_{KS}}, {\rho'}(t)] +[v'_{Hxc}(t), {\rho_0}] +[v'_{ext} (t),{\rho_0}]  
\end{equation}
where ${H^{GS}_{KS}}$ is time independent ground-state Hamiltonian, $v'_{ext}(t)$ is 
the perturbing external potential, $v'_{Hxc}(t)$ is a linear variation of the Hartree plus 
exchange-correlation potential
induced by $\rho'(t)$.  $\rho'(t)=\rho(t) - \rho_0$.
The linearized Liouvillian equation is given by
\begin{equation}
i{d {\rho'}(t)  \over dt}=L\cdot {\rho'} +[v'_{ext} (t), {\rho_0}],    
\end{equation}
The action of the Liouvillian super-operator L on $\rho'$ is given by
\begin{equation}
L\cdot {\rho'}= [{H^{GS}_{KS}}, {\rho'} ] +[v'_{Hxc}[{\rho'}] (t), {\rho_0}]   
\end{equation}
Fourier analyzing Eq. (5) we have:
\begin{equation} 
(w-L)\cdot {\rho'(w)} = [{v'_{ext}} (w), {\rho_0}]   
\end{equation}
if $v'_{ext} ({\bf r},w)=-{\bf E}(w)\cdot {\bf r}$, the response of the dipole to an external 
electric field ${\bf E}(w)$ is given by
\begin{equation}
d_i(w)=\sum \alpha_{ij} E_j (w) 
\end{equation}
The dynamical polarizability, $\alpha_{ij}(\omega)$ is defined by
\begin{equation}
\alpha_{ij} (\omega) = -<r_i|{[r_j, \rho_0] \over (w-L)}>    
\end{equation}
Eq. (9) indicates that the dynamical polarizability can be expressed as an 
appropriate off-diagonal matrix element of the resolvent of the non-Hermitian 
Liouvillian superoperator between two orthogonal vectors \cite{c13}. These matrix
elements are calculated using the Lanczos algorithm \cite{c14,c15}.  
Finally the absolute DOS's of the C$_{60}$ are obtained from 
\begin{equation}
S(\omega)={4 \omega \over 3\pi } \sum I_m \alpha_{jj} 
\end{equation}

After the C$_{60}$ calculation a Xe atom was introduced into the center of the C$_{60}$.
The DOS's of the Xe@C$_{60}$ was evaluated using the same procedure as described above. 
Finally the DOS's of Xe atom encapsulated in the C$_{60}$ were obtained by subtracting the 
DOS's of the C$_{60}$ from the corresponding DOS's of the Xe@C$_{60}$ molecule.

\section{Results}
Fig. 1 presents the spectra of a Xe atom encapsulated inside the C$_{60}$  
in the energy range of Xe 4d giant resonance.
Solid curve and dashed curve represent, respectively the results of 
the TDDFT [1] and using our C$_{60}$ model potential \cite{r6}. 
Both the TDDFT and model potential results confirm the 
three main peaks observed in the experiment \cite{d16} if the spectra are reduced by a factor of 
8 or 10 as discussed in Refs. [1,13]. 

\begin{figure}[b]
\includegraphics{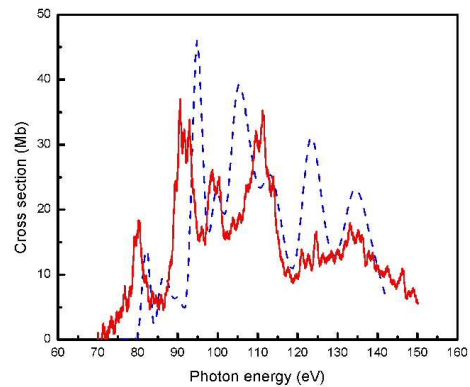}
\caption{\label{fig1} 
(Color online) Photoabsorption spectra of a Xe atom encapsulated inside C$_{60}$ 
in the energy range of the Xe 4d giant resonance.
Solid and dashed curves represent, respectively the results of our TDDFT [1] and 
the C$_{60}$ model potential calculation [13]. 
}
\end{figure}

The photoabsorption spectrum of the Xe atom encapsulated in the C$_{60}$ in the energy 
region of C$_{60}$ giant resonance can be evaluated using the same method as mentioned above.
It is well known that there are two giant 
resonances in the C$_{60}$ photoionization spectrum \cite{r17,r18,r19}, which correspond 
to a collective oscillation of 
the 240 delocalized valence electrons relative to the ionic cage of the C$_{60}$ fullerene. 
These delocalized electrons are distributed over the
surface but confined in a thickness of a single carbon atom in the radial direction.
The valence electrons can move coherently in photoexcitation.
The strong peak around 6.3 eV \cite{r17} is interpreted
as collective excitations of the $\pi$ electrons, the giant resonance
between 15 -25 eV, peaked at 22 eV belongs to the $\sigma$ electrons \cite{r18,r19}.
In addition to these there are two broad structures around respectively, 29 eV and 32 eV
,which are caused by shape resonance \cite{r19} .
The whole C$_{60}$ spectrum can be found in Ref. [1].

\begin{figure}
\includegraphics{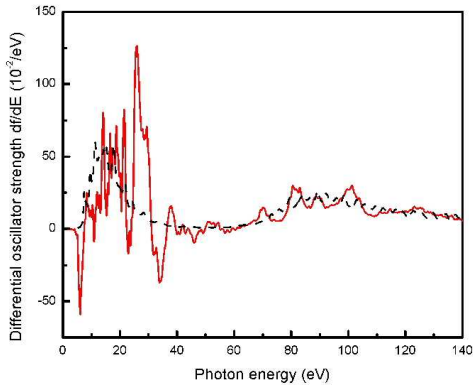}
\caption{\label{fig2}
Absolute DOS's for the photoabsorption of the Xe atom.
Solid and dashed curves represent, respectively the TDDFT results of the encapsulated  
and free Xe atoms.
}
\end{figure}

Solid and dashed curves in Fig. 2 are, 
respectively the absolute DOS's for the photoabsorption of
the encapsulated and free Xe atom calculated by TDDFT method. 
In the energy range of 65 -140 eV, the solid curve is same as that  
in Fig. 1.
Within the C$_{60}$ giant resonance region
the spectrum of the Xe atom encapsulated in the C$_{60}$ shows strong oscillation. 
It is noted that the absolute DOS's for the atom encapsulated in the C$_{60}$ 
was obtained by subtracting the DOS's of the C$_{60}$ from the corresponding results 
of the Xe@C$_{60}$ molecule. Therefore 
if the DOS's of the C$_{60}$ fullerene is larger than the DOS's of the Xe@C$_{60}$ molecule 
the curve has negative values. This means that   
the Xe atom disturbed the C$_{60}$ resonance and reduced the DOS's of the C$_{60}$. 
Fig. 2 indicates that the negative value usually happens in the low energy range. 
If the photon energy cannot ionize the Xe 5$p$ electron 
the C$_{60}$ valence electrons  
hardly transfer their resonance energy to the Xe electrons through intershell coupling.
Otherwise the energy transfer from
C$_{60}$ valence electrons to the photoelectron is strong. 
Because of this, the peak at 25 eV may be related to the Xe 5$s$ ionization (see Fig. 2). The 
ionization threshold for the 5$s$ of a free Xe is 25.4 eV. 

Strong oscillations as in Fig. 2 
can also be found in the spectra of the Ne and Ar atoms encapsulated in the C$_{60}$. 
In the following, we first evaluate the DOS's for the 
photoabsorption of the free Ne and Ar atoms.
The ultrasoft pseudopotentials for these atoms have been 
created using the PBE \cite{r10} exchange-correlation functional and 
including the relativistic effect and nonlinear core corrections \cite{v1}. 
The valence electrons for Ne and Ar are respectively, 2$s$, 2$p$ and 3$s$, 3$p$.
Solid curves in Figs. 3 and 4 are, respectively the 
calculated absolute DOS's for the free Ne and Ar atoms. 
The open circles in Figs. 3 and 4 represent the correspondent experimental data \cite{r21}. 
The measurements used the low-resolution dipole spectrometer. The data 
started from the first ionization potentials, 21.6 eV of the Ne 2$p$ state and 
16.0 eV of the Ar 3$p$ state. 
The present TDDFT calculations are in good agreement with the experimental
results reported by Ref. \cite{r21}.

The solid curves in both Figs. 3 and 4 rise rapidly initially. The experimental 
spectra of Ne and Ar, reach their maxima, respectively at the 31.5 and 21.6 eV.
After reaching its maximum 
the Ar curve drops quickly to approach the Cooper
minimum around 47 eV. The 3$s$ ionization threshold of a free Ar atom is about 34.7 eV.  
At this energy a small peak is found in the theoretical curve.

\begin{figure}
\includegraphics{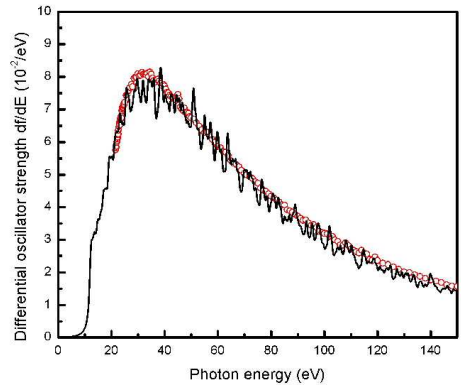}
\caption{\label{fig3} 
(Color online) Absolute DOS's for the Photoabsorption of 
a free Neon atom. Solid curve and open circles represent, respectively our TDDFT calculation
and the experimental data \cite{r21}. 
}
\end{figure}

\begin{figure}
\includegraphics{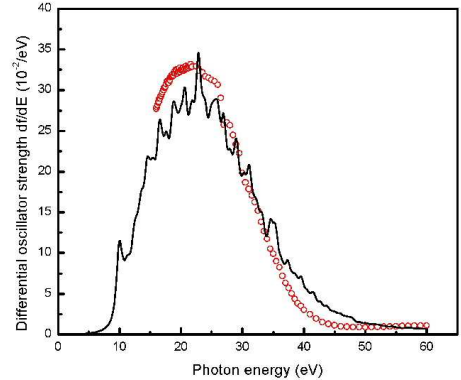}
\caption{\label{fig4}
(Color online) The symbols have the same meaning as in figure 3, except that 
the curve and open circles represent the free Ar atom.
}
\end{figure}

The curves in Figs. 3 and 4 demonstrate that our ultrasoft pseudopotentials describe 
the photoabsorption 
processes of the Ne and Ar atoms very well. The photoabsorption spectra of 
the endohedral fullerenes, Ne@C$_{60}$ and Ar@C$_{60}$ have also been calculated using the same 
TDDFT method.  The DOS's of the Ne and Ar atoms encapsulated in the C$_{60}$ are obtained by 
subtracting the DOS's
of the C$_{60}$ from the correspondent DOS's of the endohedral fullerenes. 

\begin{figure}
\includegraphics{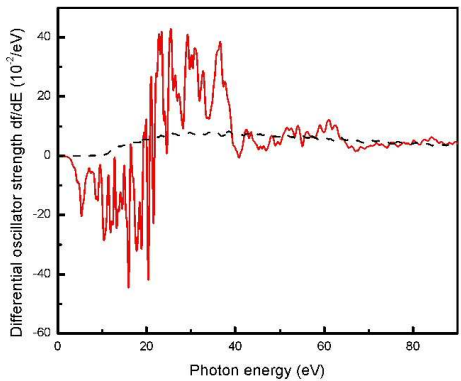}
\caption{
(Color online) Solid and dashed curves are respectively, the calculated absolute DOS's for
the photoabsorption of the encapsulated and free Ne atoms. 
}
\label{fig:5}    
\end{figure}

Dashed and solid curves in Figs. 5 and 6 represent, respectively the DOS's
for the photoabsorption of the free and encaged Ne and Ar atoms.
The solid curves in Figs. 5 and 6 demonstrate the strong oscillations in the 
region of the C$_{60}$ giant resonance. The solid curve in Fig. 5 can be  
divided roughly into two parts, a positive part and a negative part.
When the photon energy is larger than the first ionization potential of the Ne atom the DOS's 
show positive values. This implies that the C$_{60}$ valence electrons transfer resonance 
energy to the photoelectron through intershell coupling. Otherwise the resonance of the C$_{60}$ 
is perturbed and the DOS's are reduced by the Ne atom.

\begin{figure}
\includegraphics{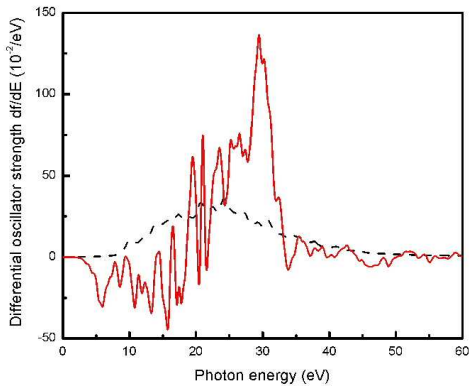}
\caption{\label{fig6}
(Color online) The symbols have the same meaning as in figure 5, except that the 
curves represent the data for the Ar atom. 
}
\end{figure}

Figure 6 shows that when the photon energy is large the DOS's are positive; otherwise
the DOS's of the C$_{60}$ are reduced by the Ar atom. This causes negative DOS's of the
Ar atom encaged in the C$_{60}$.

The solid curves in figures 2, 5 and 6 demonstrate that the DOS's of the atoms encaged in the C$_{60}$ are
positive only when the photon energies are large. This indicates that strong energy
transfer can occur only between the C$_{60}$ valence electrons and the photoelectron.   
   
\section{Conclusion}
   In conclusion, the absolute DOS's for
the photoabsorption of the Ne, Ar, and Xe atoms encapsulated in the C$_{60}$ have been calculated 
in the energy regions both inside and outside the C$_{60}$ giant resonance.
Within the C$_{60}$ giant resonance the spectra of the Ne, Ar, and Xe atoms encaged in the C$_{60}$ 
demonstrate strong oscillation with significant increase of the DOS's. 
This large enhancement of the DOS's cannot be explained by confinement resonance but is
due to the energy transfer from the C$_{60}$ valence electrons to the photoelectron  
through the intershell coupling. The energy transfer occurs 
only when the photon energy is large enough to ionize the atom encapsulated in the C$_{60}$.
To study the confinement resonance it is recommended to avoid the C$_{60}$ 
resonance region by selecting the photon energy larger than 40 eV.

\begin{acknowledgments} 
   This work was supported by the U.S. DOE, Division of Chemical Sciences,
Geosciences and Biosciences, Office of Basic Energy Sciences, Office
of Energy Research, AFOSR and Army Research Office (Grant W911NF-11-1-0194).
Calculations used Kraken System (account number TG-DMR110034) of the National 
Institute for Computational
Science, The University of Tennessee. 
\end{acknowledgments}

\end{document}